\documentclass[twocolumn]{emulateapj}

\shorttitle{Mass measurements in protoplanetary disks}
\shortauthors{McClure et al.}

\begin{document}

\title{Mass measurements in protoplanetary disks from hydrogen deuteride}

\author{M. K. McClure\altaffilmark{1}, E. A. Bergin\altaffilmark{2}, L. I. Cleeves\altaffilmark{3, 2}, E. F. van Dishoeck\altaffilmark{4}, G. A. Blake\altaffilmark{5}, N. J. Evans II\altaffilmark{6}, J. D. Green\altaffilmark{6, 7}, Th. Henning\altaffilmark{8}, K. I. {\"O}berg\altaffilmark{3}, K. M. Pontoppidan\altaffilmark{7}, C. Salyk\altaffilmark{9}}

\altaffiltext{1}{Karl-Schwarzschild-Stra{\ss}e 2, 85748 Garching bei M{\"u}nchen; mmcclure@eso.org}
\altaffiltext{2}{Department of Astronomy, The University of Michigan, 500 Church St., 830 Dennison Bldg., Ann Arbor, MI 48109; ebergin@umich.edu}
\altaffiltext{3}{Harvard-Smithsonian Center for Astrophysics, 60 Garden St, MS 16,Cambridge, MA 02138; ilse.cleeves@cfa.harvard.edu, koberg@cfa.harvard.edu}
\altaffiltext{4}{Leiden Observatory, Leiden University, P.O. Box 9513, 2300 RA Leiden, The Netherlands; ewine@strw.leidenuniv.nl}

\altaffiltext{5}{California Institute of Technology, Mail Stop 150-21, Pasadena, CA 91125; gab@gps.caltech.edu}
\altaffiltext{6}{The University of Texas at Austin, Department of Astronomy, RLM 15.312A, Austin, TX 78712; nje@astro.as.utexas.edu}
\altaffiltext{7}{Space Telescope Science Institute, 3700 San Martin Drive, Baltimore, MD 21218;  jgreen@stsci.edu, pontoppi@stsci.edu}
\altaffiltext{8}{Max-Planck-Institut f{\"u}r  Astronomie, K{\"o}nigstuhl 17, D-69117 Heidelberg; henning@mpia.de}
\altaffiltext{9}{Physics and Astronomy Department, Vassar College, Vassar College Box 745, Poughkeepsie, NY 12604; cosalyk@vassar.edu}

\begin{abstract}

The total gas mass of a protoplanetary disk is a fundamental, but poorly determined, quantity. A new technique \citep{bergin+13} has been demonstrated to assess directly the bulk molecular gas reservoir of molecular hydrogen using the HD J=1-0 line at 112 $\mu$m. In this work we present a {\it Herschel} Space Observatory\footnote{Herschel is an ESA space observatory with science instruments provided by European-led Principal Investigator consortia and with important participation from NASA.}  survey of six additional T Tauri disks in the HD line.  Line emission is detected at $>$3$\sigma$ significance in two cases: DM Tau and GM Aur.  For the other four disks, we establish upper limits to the line flux. Using detailed disk structure and ray tracing models, we calculate the temperature structure and dust mass from modeling the observed spectral energy distributions, and include the effect of UV gas heating to determine the amount of gas required to fit the HD line.  The range of gas masses are 1.0-4.7$\times10^{-2}$ for DM Tau and 2.5-20.4$\times10^{-2}$ for GM Aur. These values are larger than those found using CO for GM Aur, while the CO-derived gas mass for DM Tau is consistent with the lower end of our mass range. This suggests a CO chemical depletion from the gas phase of up to a factor of five for DM Tau and up to two orders of magnitude for GM Aur. We discuss how future analysis can narrow the mass ranges further.

\end{abstract}

\keywords{Protoplanetary disks --- radiative transfer --- astrobiology}

\section{Introduction}
\label{intro}

As a fundamental property of protoplanetary disks, the total disk gas mass is of critical importance to our understanding of disk evolution \citep{armitage+11}.  This question, in turn, informs the field of planet formation: the disk gas mass at various ages is taken as an initial condition by planetary population synthesis models \citep{mordasini+12a}.  The majority constituent of circumstellar gas, H$_2$, is difficult to observe directly at low temperatures because it lacks a dipole moment and its transitions lie at wavelengths that are difficult to observe. Instead, disk gas masses are typically inferred through one of two proxies: sub-millimeter observations of either the dusty disk component \citep{beckwith+90, aw05, aw07b} or molecular gas, i.e. CO \citep{dutrey+96, williams+14a}. 

However, both methods are inherently uncertain. The former assumes specific dust/gas mass ratios, dust opacities, and a disk radius as input, which can be affected by differing grain sizes, compositions, local dust overdensities, or a lack of spatial resolution \citep{testi+14}. The latter assumes a CO/H$_2$ abundance which is known to vary across the disk due to UV photodissociation in the surface layers and freeze-out of CO in the cold midplane \citep{dutrey+97a, vanzadelhoff+01, reboussin+15}. Moreover, isotope-selective processes need to be taken into account to properly infer disk masses from observations of C$^{18}$O lines \citep{visser+09b, miotello+14a}, and models assume an overall abundance of carbon in volatile form as input parameter. All of these uncertainties combined can lead to mass estimates differing by several orders of magnitude \citep{bruderer+12,favre+13, bergin+14a,kama+16b}. A third method has been demonstrated by \citet{bergin+13} to assess directly the bulk molecular gas reservoir of molecular hydrogen using an isotopologue of H$_2$, hydrogen deuteride (HD); compared with the previously discussed proxies, HD should have a constant abundance relative to H$_2$ throughout the disk. Taking advantage of the wavelength coverage and sensitivity of the {\it Herschel} Space Observatory, these authors were able to observe the 1-0 transition of HD at 112$\mu$m in the nearest protoplanetary disk, TW Hya.  Based on the line strength and detailed chemical modeling, they suggested a lower limit to the disk gas mass for TW Hya of 0.06 M$_{\odot}$ or six times the `minimum mass solar nebula' \citep[MMSN=0.01 M$_{\odot}$,][]{hayashi+81}. That work, however, only encompassed the first object of a larger sample observed with {\it Herschel}.  

We present the analysis of the remaining sample of six T Tauri stars, the selection of which is described in Section \ref{obsred}. Through an approach combining disk structure, UV radiation, and gas heating/cooling models, we determine a set of dust and gas density and temperature structures that fit the spectral energy distribution (SED) and HD line emission (Section \ref{allanalysis}). We find a range of disk masses that fit the observations for the two sources with HD detections (DM Tau and GM Aur, Section \ref{results}). In Section \ref{discussion}, we compare our results to CO mass measurements and discuss how the uncertainties in mass might be reduced further.

\begin{deluxetable*}{ccccccc}
\tabletypesize{\small}   
\tablewidth{0pt}
\tablecaption{Observations and Stellar Parameters}
\tablehead{
\colhead{Parameter} & 
          \colhead{DM Tau} & \colhead{GM Aur} & \colhead{VZ Cha} & \colhead{AA Tau} & \colhead{FZ Tau} & \colhead{LkCa 15}  }
\startdata
OBSID & 1342239747 & 1342243524 & 1342232613 & 1342239749 & 1342239750 & 1342240148 \\
Date & 2012-02-26 & 2012-03-25 & 2011-11-22	 & 2012-02-27 & 2012-02-27 & 2012-02-17 \\
RA (J2000)$^a$ & 04 33 48.718  & 04 55 10.983  & 11 09 23.790 & 04 34 55.424  & 04 32 31.764  & 04 39 17.796  \\
Dec. (J2000)$^a$ & +18 10 09.99 & +30 21 59.54  & -76 23 20.76  & +24 28 53.16 & +24 20 03.00 & +22 21 03.48  \\
$d$ (pc) 			& 140    & 140   & 160 & 140 & 140 & 140 \\
$T_{eff}$ (K) 		& 3720 & 4350 & 3780 & 4060 & 3850 & 4730 \\
$A_V$ (mag) 		& 0.5    & 0.8    & 1.9 & 1.3 & 6.6 &  1.7 \\
$M_*$ (M$_{\odot}$) & 0.65  & 1.1    & 0.85 & 0.8 & 0.6 & 1.3 \\
$R_*$ (R$_{\odot}$) & 1.2     & 1.7    & 1.6 & 1.8 & 2.3 &  1.6 \\
$\dot{M}$ (M$_{\odot}yr^{-1}$) &  $2\times10^{-9}$ & 4.7$\times$10$^{-9}$ & 6.2$\times$10$^{-8}$ & 6$\times$10$^{-9}$ & 3.4$\times$10$^{-7}$ & 3.3$\times$10$^{-9}$\\
$i$ (\degr)			& 35      & 55     & 60$^b$ & 71 & 75 & 42\\
$R_{out}$ (AU)	 	& 160    & 300  & 140$^c$ & 140  & 140$^c$ & 300

\enddata
\label{obstab}
\tablecomments{$^a$ RA and Dec are given in (h m s) and (deg arcmin arcsec), respectively, $^b$ VZ Cha does not have a measured inclination, so we assume an average value. $^c$VZ Cha and FZ Tau do not have measured R$_{out}$, so we assume a value of 140 AU.  References for stellar parameters and disk geometry: DM Tau \citep{calvet+05,teague+15, isella+09}, GM Aur \citep{espaillat+11}, VZ Cha \citep{manara+16a}, AA Tau \citep{mcclure+15, cox+13}, FZ Tau \citep{ricci+10a}, 
LkCa15 \citep{espaillat+10}}
\end{deluxetable*}

\section{Observations and data reduction}
\label{obsred}

For these pioneering observations of this line there was strong concern about its detectability due to the fact that the dust continuum of many disks is strong and optically thick at 112 $\mu$m, and the predicted line/continuum ratio was low.  Given this issue, the sample was selected from objects that had previous molecular line detections, and all of our targets were selected to have weak continuum emission (below 5 Jy) at 100 $\mu$m. As an additional factor, our source selection is weighted towards highly settled systems where the dust photosphere is likely below the main layers of gas emission or systems where there is evidence for substantial grain growth (e.g. transition disks); both would lower the 100 $\mu$m dust optical depth.  This can be tracked by the $n_{13-31}$ index, a spectral slope measured between 13 and 31$\mu$m by the {\it Spitzer} InfraRed Spectrograph \citep{furlan+06}.  However, since the goal of this program was to detect a line, we chose deeper integrations on sources selected to maximize detection rather than conducting a shallow survey of a sample that tests a range of parameter space.

The sample was observed by {\it Herschel} \citep{pilbratt+10} with the Photodetector Array Camera and Spectrometer \citep[PACS;][]{poglitsch+10} through program OT1\textunderscore ebergin\textunderscore 4 (PI: Bergin) in range spectroscopy mode for 8320 seconds per target.  The observation IDs are given in Table \ref{obstab}.  PACS is a 5$\times$5 array of 9\farcs4$\times$9\farcs4 spatial pixels `spaxels' at spectral resolution R $\sim$ 1500-3000.  The nominal pointing RMS of the telescope is 2''.  The data were reduced using the HIPE interactive pipeline version 13 / CalTree 65, provided by the Herschel Science Center, the most current version at the time of reprocessing during summer 2014.  The linescan mode has too narrow wavelength coverage (less than 2 $\mu$m width) to utilize the pipeline `jitter' correction, but other corrections are included.  A complete description of the pipeline, along with estimates of precision and efficiency, can be found in \citet{green+16}. The spectra are rebinned by a factor of two before post-processing.

\begin{figure*}
\centering{\includegraphics[angle=0, scale=0.95]{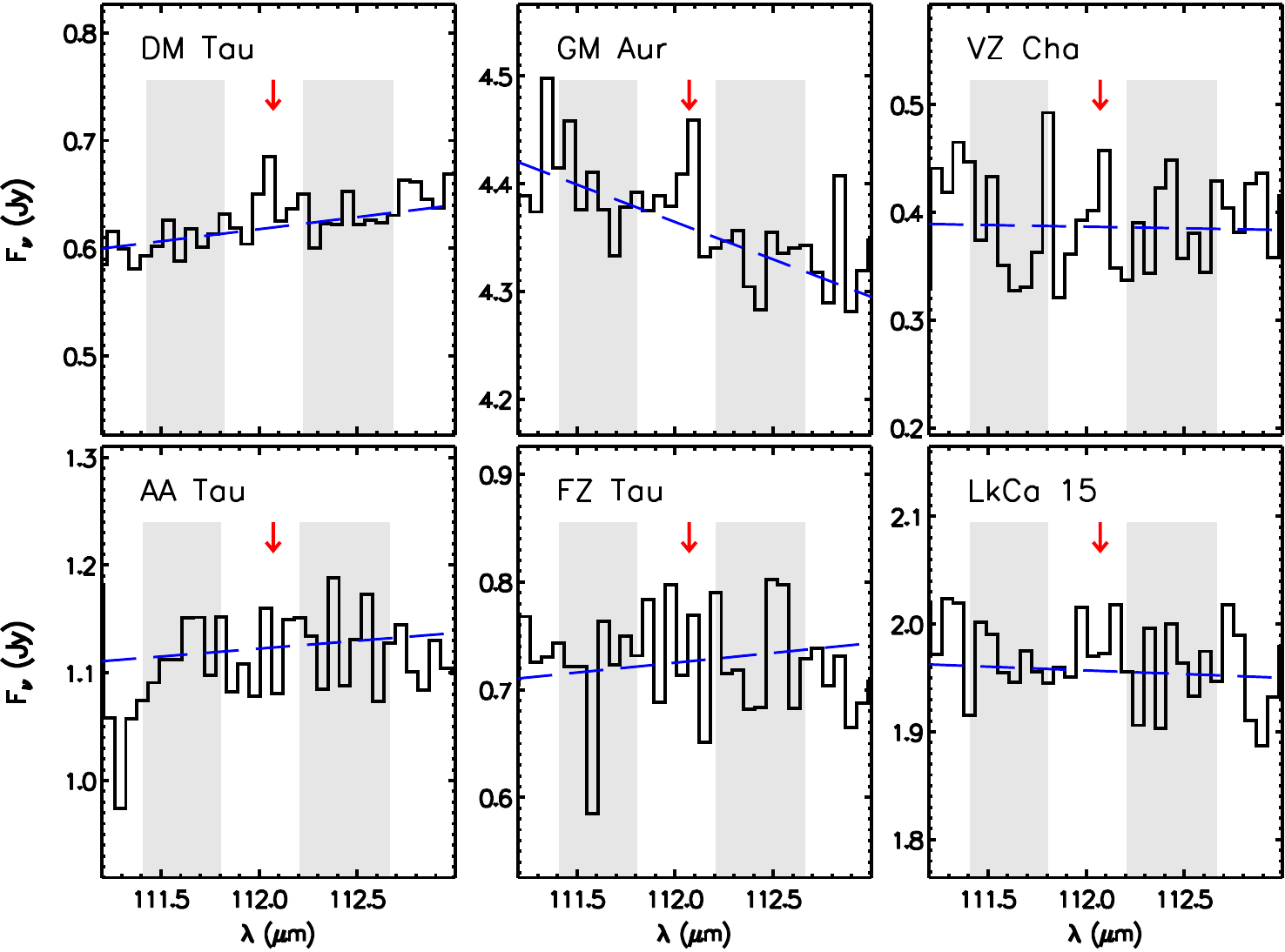}}
\caption{PACS spectra of our sample in the region around the HD line. The continuum (blue, dashed line) was determined by fitting a first order polynomial to regions on either side of the feature (grey fill). The red arrow indicates the location of the HD line. \label{rawspec}}
\end{figure*}

\begin{figure*}
\centering{\includegraphics[angle=0, scale=0.95]{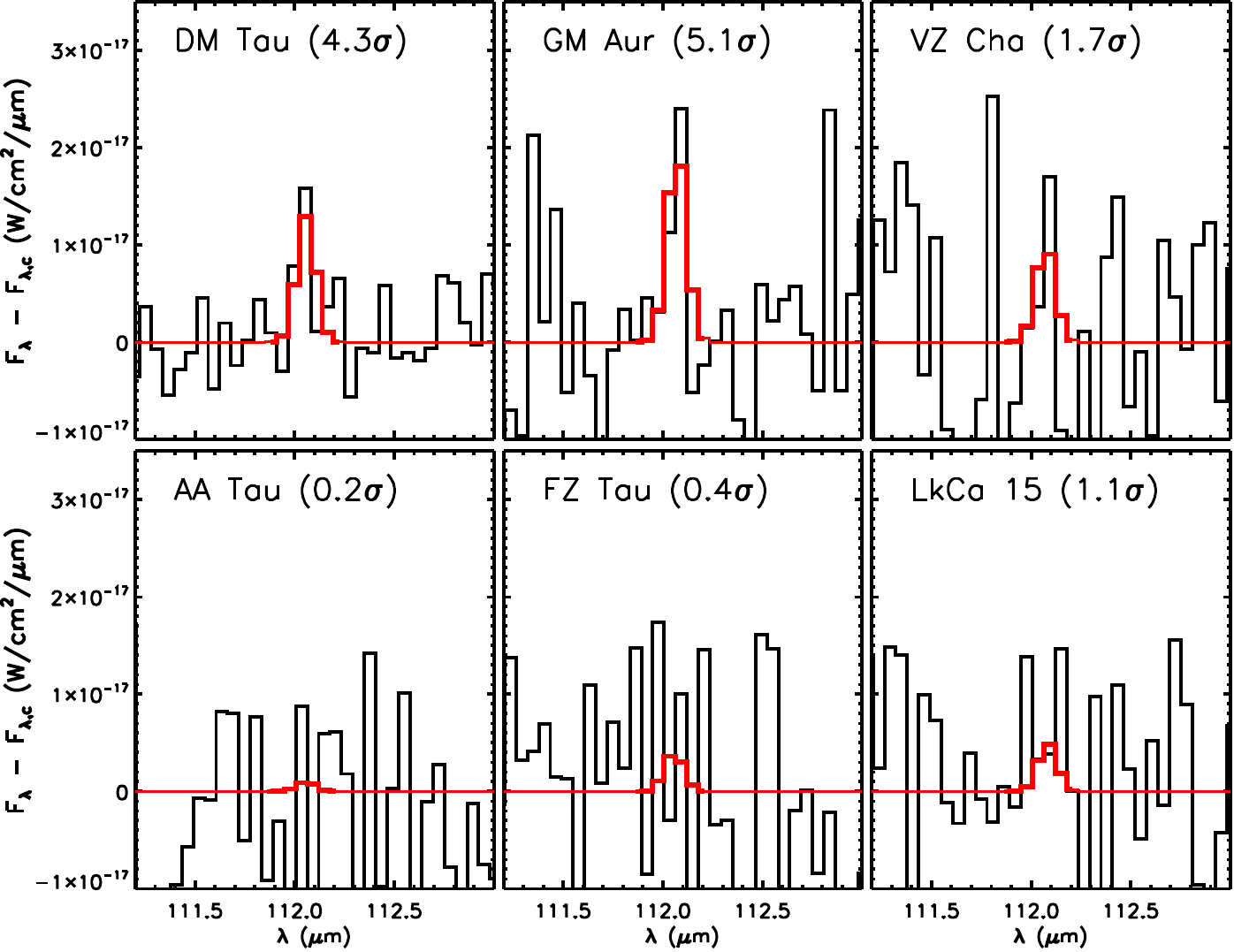}}
\caption{Gaussian line fits (red) to the continuum subtracted spectra (black). The FWHM and central wavelength were fixed at the instrumental and theoretical values, respectively. The thicker red line indicates the region over which the RMS deviation of the residuals is measured. The 1$\sigma$ uncertainty calculation is described in Section \ref{obsred}. \label{normspec}}
\end{figure*}

In Figure \ref{rawspec} we display the reduced spectra of each target, along with a first-order polynomial continuum fit to regions on either side of the HD line.  After subtracting the continuum, we measured the emission through fits to a Gaussian line profile (Figure \ref{normspec}).  The full width at half maximum (FWHM) is fixed at 0.115 $\mu$m, according to the instrument profile, while the central wavelength, $\lambda_c$, is initially set at 112.072$\mu$m (CDMS linelist) but allowed to vary within 0.01$\mu$m to account for uncertainty in the wavelength calibration.  We obtain a best fit to the peak value at the central wavelength position, $F_0$, and measure the uncertainty in the value by the root mean squared (RMS) value of the fit residuals over the line profile. The integrated flux, $F_{int}$, is taken over the six-channel width of the best-fitting profile, and its uncertainty calculated from the FWHM, $F_{0}$, and the RMS uncertainty. Two of the six targets, DM Tau and GM Aur, show weak but statistically significant HD emission. The peak and integrated fluxes are reported in Table \ref{fluxtab}. For the four disks in which HD was not detected, the listed integrated fluxes are 3$\sigma$ upper limits.  

\section{Analysis}
\label{allanalysis}

The amount of emission in the HD line depends on the fractional population of the HD J=1 level, which is a function of both the gas temperature and the total disk gas mass. To determine the best fitting disk mass, while taking into account simultaneously the effects of gas temperature, we use three modeling prescriptions to fit the combination of the observed spectral energy distribution (SED; dominated by details of dust structure) and HD gas line. First, we fit the SED with the \citet{dalessio+06} disk structure models to determine the dust density and temperature at each position ($\rho_{dust}$, $T_{dust}$, $M_{dust}$) and produce a grid of initial gas density distributions and masses ($\rho_{gas}$, $M_{gas}$) to be used in determining the gas temperature and disk gas masses (Section \ref{dalessio_mod}). The densities and temperature structure are used as input to the \citet{bethellbergin09} continuum radiative transfer code to calculate the UV flux at each point in the disk (Section \ref{gastemp}). This information is used with the prescription of \citet{bruderer+13} to calculate a separate gas temperature structure, $T_{gas}$. The separate density and temperature structures for the dust and gas are then given as input to RADLite \citep{pontoppidan+09}, \textbf{with an assumed abundance of $x_{HD}$=3$\times$10$^{-5}$ relative to H$_2$ (Section \ref{radlite})}.  The line emission produced by RADLite is compared with the observed data, and, for the case of the disks with detections, we iterate this procedure over a range of disk temperature structures and gas masses until the line emission fits. We give more detailed descriptions of input to each code below.

 \begin{deluxetable}{cccc}
\tabletypesize{\small}   
\tablewidth{0pt}
\tablecaption{Measured line fluxes}
\tablehead{\colhead{Star} & \colhead{F$_{0}$} &\colhead{F$_{int}$}  & \colhead{S/N}  \\
   &   ($\times$10$^{-17}$  & ($\times$10$^{-18}$  &   \\
   &   W m$^{-2}$ $\mu$m$^{-1}$)  & W m$^{-2}$)  &    }
\startdata
DM Tau & 1.30$\pm$0.43 & 1.6$\pm$0.4 & 4.3 \\
GM Aur &  2.00$\pm$0.55 & 2.5$\pm$0.5 & 5.1 \\
VZ Cha & 1.00$\pm$0.81 & $<$2.1$^a$ & $-$ \\
AA Tau  &  0.10$\pm$0.85 & $<$2.2$^a$ & $-$ \\
FZ Tau &  0.40$\pm$1.40 & $<$3.6$^a$ & $-$ \\
LkCa 15 &  0.50$\pm$0.66 & $<$1.7$^a$ & $-$  
\enddata
\label{fluxtab}
\tablecomments{Each flux was measured using a gaussian fit with a fixed FWHM of 0.115$\mu$m and central wavelength of 112.07159$\mu$m.  Column 2: Peak flux at central wavelength, with residual RMS uncertainty, Column 3: Integrated flux, $^a$: 3$\sigma$ upper limit.}
\end{deluxetable}

\subsection{Initial disk structure model}
\label{dalessio_mod}
To construct the initial disk temperature and density structures, we used the updated D'Alessio Irradiated Accretion Disk (DIAD) 1+1D disk structure code  \citep{dalessio+04, dalessio+06}, which calculates the structure equations for a gas- and dust- rich disk assuming hydrostatic equilibrium. Heating is provided by both viscous accretion and irradiation by the star and its accretion shock, and it is assumed that the dust and gas temperatures are coupled throughout the disk. This assumption is reasonable for the cooler molecular layers we expect to probe with HD, but ultimately we relax this requirement at a later stage to test this assumption (see Section \ref{gastemp}). The dust in the disk is distributed between two populations: one in the disk midplane (i.e. below 0.1$H$, where $H$ is the gas pressure scale height of the disk) and the other in the disk upper layers (above 0.1$H$). The distribution of dust between these layers sets the temperature structure, while the total gas mass is a function of the surface density, disk size, and dust/gas mass ratio.  Since we use a physical model for the disk surface density that enforces hydrostatic equilibrium self-consistently, the dust/gas mass ratio is an input rather than the disk gas mass, and it is not possible to vary the disk temperature and gas mass entirely independently of each other. The input parameters that affect the temperature structure and mass are detailed below.

\subsubsection{Dust/gas mass ratio and dust temperature structure}
\label{dust/gas_ratio}
The disk temperature structure is strongly affected by the settling of dust grains from the disk upper layers to the midplane, as a reduction in small grains in the upper layers allows stellar radiation to penetrate deeper into the disk. This effect is particularly pronounced for the separate UV gas heating module (Section \ref{gastemp}), as discussed in Section \ref{depletion_uv}. The amount of dust depletion is measured relative to the `standard' dust/gas mass ratio: $\epsilon_{small}=\chi_{upper layers}/\chi_{standard}$, where $\chi_{standard}$ is determined from the dust species mass fractions. Larger values of $\epsilon_{small}$ indicate less dust depletion, e.g. $\epsilon_{small}$=0.5 indicates 50\% of the dust is depleted, while a value of 0.01 indicates that 99\% is depleted. The dust/gas ratio in the midplane is enhanced by the solids that settle out of the upper layers. The value of dust enhancement, described by $\epsilon_{big}=\chi_{midplane}/\chi_{standard}$, is coupled to $\epsilon_{small}$ so as to conserve mass vertically at a given radius \citep[see Table 3 of][]{dalessio+06}.

The dust in our model has a power law size distribution, from a minimum size of 0.005$\mu$m to a maximum size $a_{max}$ with a power of -3.5 \citep{mrn77}. In the midplane, $a_{max}$ is fixed at 1mm. In the upper layers, it is fixed at the values found by \citet{espaillat+11} for DM Tau, GM Aur, and LkCa 15 and allowed to vary for the other disks. The dust grain species and opacities differ from \citet{dalessio+06}. For five of the six disks, we use only silicates and graphite. The silicate opacities are taken to be those that best fit the {\it Spitzer} IRS spectra in \citet{sargent+09b}: amorphous, glassy olivine and pyroxene \citep{jaeger+94,dorschner+95} or crystalline forsterite and enstatite \citep{chihara+02, sogawa+06} with a mass fraction of 0.004 relative to the gas. The opacity and mass fraction of 0.0025 for graphite is taken from \citet{dl84}. For these disks $\chi_{standard}$=0.0065. Water ice is included only for AA Tau, in which it has been detected through spectral features \citep{chiang+01,mcclure+15}. In that case the water ice mass fraction is 0.002, $\chi_{standard}$=0.0085 and the composite ice opacities are taken directly from \citep{mcclure+15}.

\subsubsection{Surface density}
\label{surface_density}
In addition to changing the temperature structure, varying the surface density, $\Sigma$, also changes $M_{gas}$, the disk gas mass. The surface density of the disk is proportional to the mass accretion rate divided by the \citet{shakura_sunyaev73} $\alpha$ parameter (which describes the disk viscosity): $\Sigma\propto\dot{M}/\alpha$.  The dust emission is directly related to the temperature, opacity, and mass of dust, where the dust emission becomes optically thin at mm wavelengths: $F_{\nu}\propto T_{dust}\kappa_{\nu}M_{dust}$. These physical quantities are related to the model input parameters: $T_{dust}$ depends on $\epsilon_{small}$, $\kappa_{\nu}$ is a function of the assumed dust species and dust/gas mass ratio, and $M_{dust}$ is determined by integrating $\Sigma_{dust}$ from the inner to outer disk radii.

By taking constraints from observations on $\dot{M}$ and $R_{out}$ (as discussed in Section \ref{observables}), using the \citet{mcclure+13b} prescription for a two layer wall to fit $R_{in}$, and fixing the dust properties, we can reduce the variable input parameters to the dust/gas ratio in the upper layers ($\epsilon_{small}$), $\alpha$, and the dust/gas at the midplane ($\epsilon_{big}$). These parameters produce a family of models that fit the (sub)millimeter photometry with the same total dust mass, $M_{dust}$, but with a range of values for the gas mass, $M_{gas}$.

\subsubsection{Observables}
\label{observables}
Certain parameters were fixed to observed values, which are given in Table \ref{obstab}. The mass accretion rate, $\dot{M}$ is taken to be that of the disk onto the star and is assumed to be constant throughout the disk. We do not account for observed variability about an average value for $\dot{M}$, which can be almost an order of magnitude over several months in the same object \citep[e.g. GM Aur,][]{ingleby+15}. However, because of the relationship between $\Sigma$, $\dot{M}$, and $\alpha$, for the same disk structure $\alpha$ could be varied to compensate for any change in the average $\dot{M}$. 

Where observations are available, we take the disk outer radius, $R_{out}$, to be that of the observed millimeter grains, rather than the larger radius defined by gas observations \citep{panic+09a,andrews+12}. It is not clear what effect the removal of large dust grains from the outer disk would have. Over time, removal of the largest grains could lower the concentration of small grains in the disk upper layers, cooling the disk beyond the radius defined by millimeter grains. Since our model does not allow for different gas and dust radii, we truncate the disk structures at the millimeter grain radius to avoid introducing an artificial contribution to the HD emission from the (mostly dust-free, by mass) regions beyond this point. This does not affect our SED fitting, which depends primarily on the dust distribution. For the HD emission, \citet{bergin+13} found for TW Hya that 90\% of the HD emission originated within 100 AU, so although there is gas outside of the millimeter grain radius, it should not contribute substantially to the line emission.

\subsection{Gas temperature}

To decouple the gas temperature, $T_{gas}$, from the dust temperature, $T_{dust}$, we need to account for heating by UV radiation.  To compute the UV radiation field at each point in the disk, we ran the \citet{bethellbergin09} code on the best-fitting 1+1D disk structures.  We assumed an input stellar UV spectrum of TW Hya \citep[$L_{FUV}$=1.86$\times$10$^{-3}$ L$_{\odot}$][]{bergin+13}, scaled according to the UV luminosity of each target, as given in Table \ref{uvtab}.  Where it was missing, we calculated the UV luminosity from the mass accretion rate or accretion luminosity, assuming the relationship between $L_{acc}$ and $L_{FUV}$ given by \citet{yang+12}.  The UV radiation field was combined with the disk gas density to calculate $\Delta T_{gas}$ according to a prescription described in the appendix of \citet{bruderer+13}, which is calibrated from detailed thermochemical models that are introduced in \citet{bruderer+12}.  We compare the differences in results obtained with $T_{gas}=T_{dust}$ and $T_{gas}>T_{dust}$ in Section \ref{results}.
\label{gastemp}

\subsection{Gas line emission}
\label{radlite}
The dual temperature and density structures, $T_{gas}$, $\rho_{gas}$, $T_{dust}$, $\rho_{dust}$ and the dust opacities were passed into the RADLite line radiative transfer code \citep{pontoppidan+09}.  We took a constant HD abundance throughout the disk of $x_{HD}$=3$\times$10$^{-5}$ relative to H$_2$, assuming HD/H$_2$ has twice the D/H value of $x_D$=1.5$\times$10$^{-5}$ within the nearest 100 pc \citep{linsky98}, as suggested by \citet{bergin+13}. As for the case of TW Hya, a lower HD abundance than that assumed here would imply larger final disk masses. Since molecular hydrogen does not freeze out, and there are relatively few ways to preferentially sequester it in other molecules or on grains on disk-wide scales, the constant abundance assumption is reasonable. 

RADLite then calculates how the HD energy levels are populated relative to the total amount of HD at each point, assuming LTE excitation.  During the ray tracing, RADLite also takes into account the dust optical depth when calculating the emergent line emission. The output spectra were resampled to the PACS instrument resolution (300 km s$^{-1}$ at 112 $\mu$m) and integrated over the emission line to be compared with the data.  For the two disks with HD detections, we compared the line fluxes calculated for several temperature structures and values of $M_{gas}$ with the observed line fluxes to determine the best-fitting gas masses.

\begin{deluxetable}{cccc}
\tabletypesize{\small}   
\tablewidth{0pt}
\tablecaption{Sample UV properties}
\tablehead{\colhead{Star} &  \colhead{$L_{acc}$} & \colhead{$L_{FUV}$} & \colhead{Ref.} \\
  &   (L$_{\odot}$) & (L$_{\odot}$)  &   }
\startdata
DM Tau &  -  & 4.02$\times$10$^{-3}$  &  1 \\
GM Aur & - & 1.85$\times$10$^{-3}$  &  1 \\ 
VZ Cha & -  & -  &  -  \\ 
AA Tau &   -  & 1.99$\times$10$^{-3}$   &  1 \\
FZ Tau &  4.64$\times$10$^{-1}$  & 1.10$\times$10$^{-2}$ & 3  \\
LkCa 15 & 2.51$\times$10$^{-2}$ & 9.83$\times$10$^{-4}$ & 2
\enddata
\label{uvtab}
\tablecomments{References: (1) \citet{yang+12}, Table 4; (2) \citet{yang+12}, Table 2; (3) \citet{ricci+10a}, from $\dot{M}$=7.2$\times$10$^{-8}$ $M_{\odot}/yr$. For comparison, \citet{bergin+13} used $L_{FUV}$=1.86$\times$10$^{-3}$ L$_{\odot}$ for TW Hya.}
\end{deluxetable}

\begin{deluxetable*}{ccccccc}
\tabletypesize{\small}   
\tablewidth{0pt}
\tablecaption{SED Model Fits}
\tablehead{
\colhead{Parameter} & 
          \colhead{DM Tau} & \colhead{GM Aur} &
          \colhead{VZ Cha} & \colhead{AA Tau} & \colhead{FZ Tau} & \colhead{LkCa 15}   }
\startdata
Sublimation wall \\
 \hline \\
$T_{in}$                        &   -   &  1200        &    1600,1000  &   1600, 750   &   1600, 1000   &   1600, 900   \\
$R_{in}$ (AU)               &   -  &  0.22          &   0.13, 0.45    &    0.12, 0.32   &   0.26, 64      &   0.10, 0.46    \\
$a_{max}$ ($\mu$m)   &   -   &  0.25         &   1, 0.25          &   1, 5             &    5, 2            &    1, 0.25   \\
silicate                          &   -  &    50\% oli  &    80\% oli       &   100\% pyr   &   100\% oli    &   100\% oli   \\
composition                  &   -  &   50\% pyr  &    20\% forst   &                      &                      &    \\
 \hline \\
Outer wall  \\
 \hline \\
 $T_{wall}$                    &   215   &  120  &     -    &   -    &   -   &  120  \\
 $R_{wall}$ (AU)           &   3        &  23   &    -    &    -   &   -   &  39  \\
 $a_{max}$ ($\mu$m)   &   1        &  1  &    -    &   -   &    -   &  0.25  \\
 silicate                         &   66\% oli,   &   50\% oli  &    -    &   -  &   -   &   100\% oli    \\
 composition                 &   34\% pyr   &   50\% pyr   &     &    &    &      \\
 \hline \\
 Disk \\
 \hline \\
 $a_{max, upper}$ ($\mu$m)   &   1         &   3         &    3  &   0.25   &   0.25   &  0.25  \\
 $a_{max, midplane}$ (mm)   &   1       &   1          &    1   &   1    &   1   &  1  \\
 m.f.$_{silicates}$                   &    0.004    &  0.004  &    0.004   &   0.004   &    0.004   &  0.004  \\
 m.f.$_{graphite}$                   &   0.0025   &  0.0025  &   0.0025   &   0.0025   &    0.0025   &  0.0025  \\
 m.f.$_{H_2O ice}$                &   1$\times$10$^{-5}$   &   1$\times$10$^{-5}$   &    1$\times$10$^{-5}$   &   0.002   &   1$\times$10$^{-5}$   &   1$\times$10$^{-5}$  \\
(Case 1, SED only) \\
$\alpha$         &  0.001  &  0.001  &  0.1  &  0.003  &  0.2  &  0.001  \\
$\epsilon_{small}$     &  0.05  &  0.05  &   0.001    &    0.01    &   0.001   &  0.001  \\
$M_{tot, SED}$ (M$_{\odot}$)   &  4.78$\times$10$^{-2}$  &  1.796$\times$10$^{-1}$  &    9.52$\times$10$^{-3}$    &    3.06$\times$10$^{-2}$   &   2.86$\times$10$^{-2}$   &   6.20$\times$10$^{-2}$  
\enddata
\label{modtab}
\tablecomments{The assumed stellar and accretion properties and disk geometries are given in Table \ref{obstab}. Dust component are: (oli)vine, (pyr)oxene, (forst)erite. The abbreviation `m.f.' indicates the mass fraction of a particular grain species relative to the gas. Total masses are given for the dust/gas ratio case where the vertical dust mass was conserved at each radius. The inner silicate sublimation rim is fit with the two-layer model description given by \citet{mcclure+13b}, and we refer the reader to this work for more details. $M_{tot}$ is defined as the sum of the gas and dust masses.}
\end{deluxetable*}

\section{Results}
\label{results}

To initiate the modeling sequence, we take the following approach. First we fit the dust SEDs of all six disks using models for which mass is conserved in a vertical column, which we call Case 1. These models allowed us to determine the disk dust mass and an acceptable range of temperature structures. We explored the following parameter space: 1$\ge \epsilon_{small} \ge$ 0.001, 0.1$\ge \alpha \ge$ 0.0001. Since mass is conserved vertically, $\epsilon_{big}$ is tied directly to $\epsilon_{small}$ and does not freely vary. The fit to the submillimeter photometry produces a best-fitting value for $\alpha$, while the fit to the mid- and far-infrared region determines the dust depletion in the disk upper layers, $\epsilon_{small}$, and the maximum dust grain size in the disk upper layers is constrained by the mid-infrared spectral features. For the four disks with non-detections, we continue the analysis with a single Case 1 model with a value for $\epsilon_{small}$ that best-fits the SED, while for DM Tau and GM Aur we take an additional family of Case 1 models with a range of $\epsilon_{small}$ about the best fitting value to test the impact on the HD line emission of varying the temperature structure. All Case 1 models for the six disks are carried forward through the gas temperature and HD line flux calculations in Sections \ref{gastemp} and \ref{radlite}.

For the best-fitting range of values of $\epsilon_{small}$ for DM Tau and GM Aur, we then vary the disk surface density via the $\alpha$ parameter to explicitly test the acceptable range of gas masses, which we call Case 2. This required fixing the disk dust mass, to maintain the fit to the (sub)millimeter photometry, by enhancing the dust/gas ratio at the disk midplane via the $\epsilon_{big}$ parameter. Doing so relaxes the mass conservation requirement assumed in Case 1, as $\epsilon_{big}$ is allowed to vary independently from $\epsilon_{small}$, with values between 12.5 and 100. Each decrease in the surface density corresponds to an increase in $\alpha$ and an increase in the midplane dust concentration to maintain the fit to the submillimeter photometry. A local enhancement of dust at the midplane is not unphysical, as dynamical processes such as radial drift or dust filtration at the edges of gaps \citep{testi+14,zhu+12} could enhance the dust concentration in the midplane interior to the outer radius defined by the millimeter sized grains. Then for each model in the Case 1 temperature grid, there is a family of Case 2 models with decreasing gas mass. We carry forward all of the Case 2 models through the gas temperature and HD line flux calculations. Combined with the Case 1 models for these disks, we can then use the HD line emission to select the best-fitting range of gas masses.

Below we discuss how varying the dust depletion, gas heating, and optical depth affects the disk structures, line emission, and disk gas masses, using DM Tau and GM Aur as examples. The best-fitting values found for these parameters are given in Table \ref{modtab} for all six disks and the SED fits for the two disks with detections are shown in Figure \ref{allseds}. The SEDs of the non-detections are given in Figure \ref{allseds} and discussed in Section \ref{nodet}.

\subsection{Case 1: Impact of disk temperature structure on HD line emission}
\label{depletion_uv}

\begin{figure*}
\centering{\includegraphics[angle=0, scale=0.45]{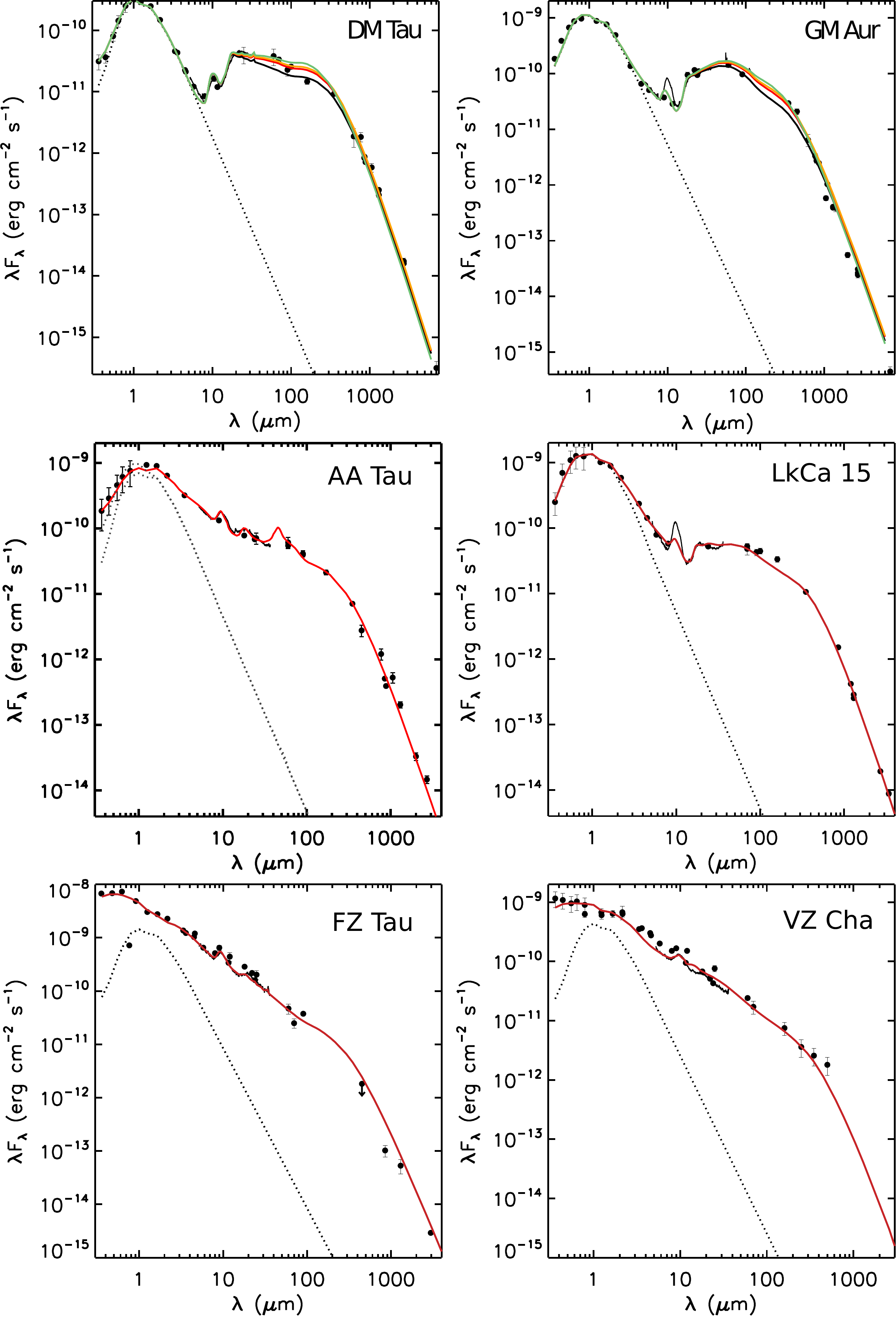}}
\caption{Best-fitting 1$+$1D models to the spectral energy distributions of our sample. Optimized values for input parameters are given in Table \ref{modtab}.  For the two disks with HD detections, DM Tau and GM Aur, we indicate the effect of varying the degree of dust depletion from the disk upper layers, $\epsilon_{small}$, on the SED fit. For these two disks, models are $\epsilon_{small}$=0.5(green), 0.1(orange), 0.05(red), and 0.01 (black). References for the photometry are: \citet{kh95}, \citet{herbig_bell_v3}, 2MASS \citep{cutri+03}, DENIS, \citet{carpenter+02}, WISE \citep{cutri+12}, AKARI IRC \citep{ishihara+10}, {\it Spitzer} IRAC \citep[][, Spitzer Enhanced Imaging Products]{luhman+08}, \citet{grafe+11}, PACS \citep{winston+12,howard+13},  AKARI FIS, {\it Spitzer} MIPS (Spitzer Enhanced Imaging Products), IRAS \citep[FSC, PSC][respectively]{moshir+90,iras_psc}, \citet{adams+90},\citet{aw05}, \citet{aw07a},\citet{aw08}, \citet{andrews+11a},\citet{bs91}, \citet{beckwith+90}, \citet{dutrey+96}, \citet{dutrey+98}, \citet{duvert+00},\citet{kitamura+02}, \citet{guilloteau+11}, \citet{isella+09},\citet{osterloh+95}, \citet{ricci+10a},\citet{rodmann+06}, \citet{weintraub+89a}.}
\label{allseds}
\end{figure*}

\begin{figure*}
\centering{\includegraphics[angle=0, scale=0.45]{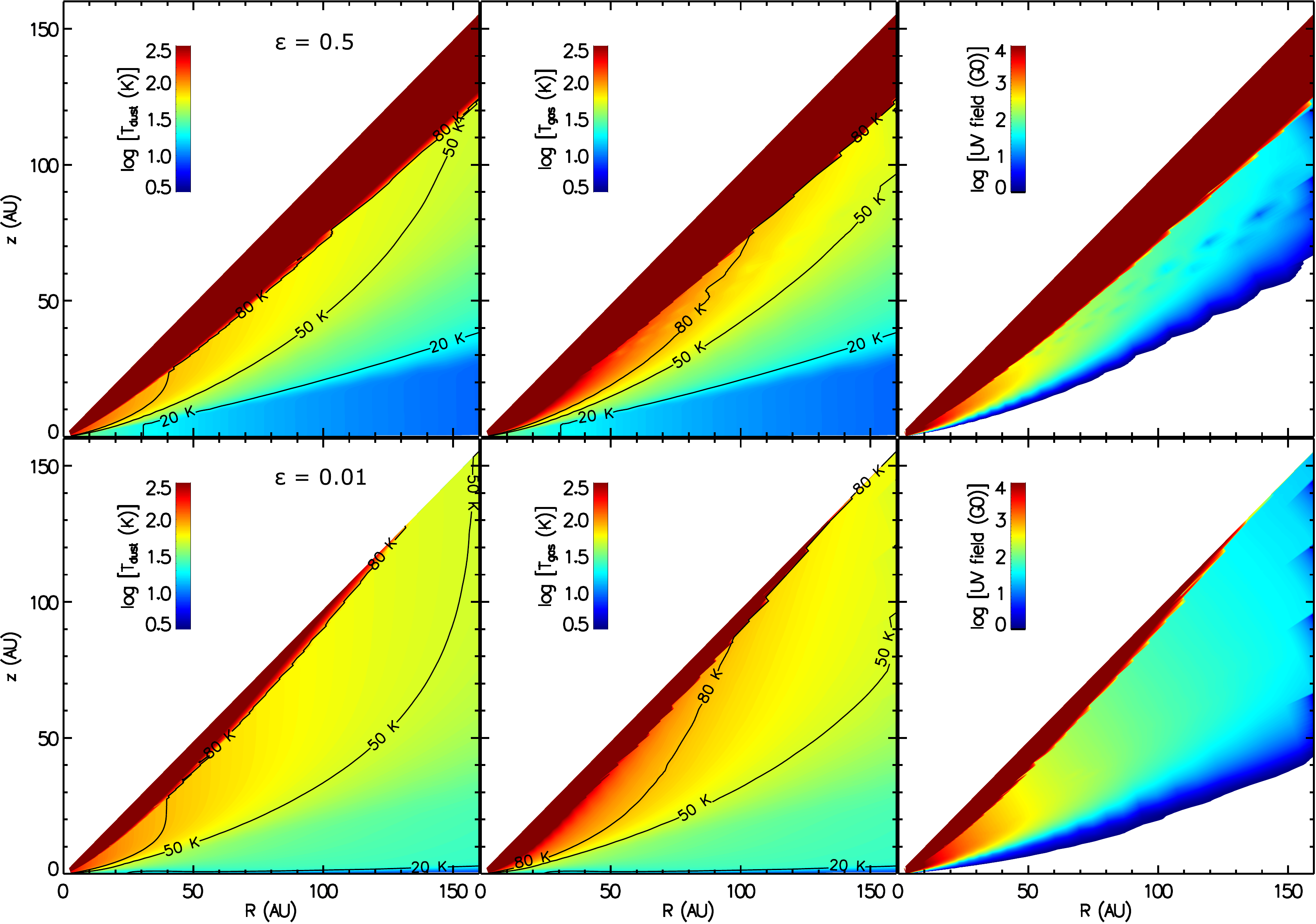}}
\caption{DM Tau disk structures. {\it Upper row:} dust and gas temperature and UV flux structures for $\epsilon_{small}$=0.5 (50\% less dust). {\it Lower row:} dust and gas temperature and UV flux structures for $\epsilon_{small}$=0.01 (99\% less dust). The UV radiation field is given in units of G0, or 1 Habing unit, equal to 1.6$\times$10$^{-3}$ erg s$^{-1}$cm$^{-2}$. \label{dmtstruct}}
\end{figure*}

\begin{figure*}
\centering{\includegraphics[angle=0, scale=0.45]{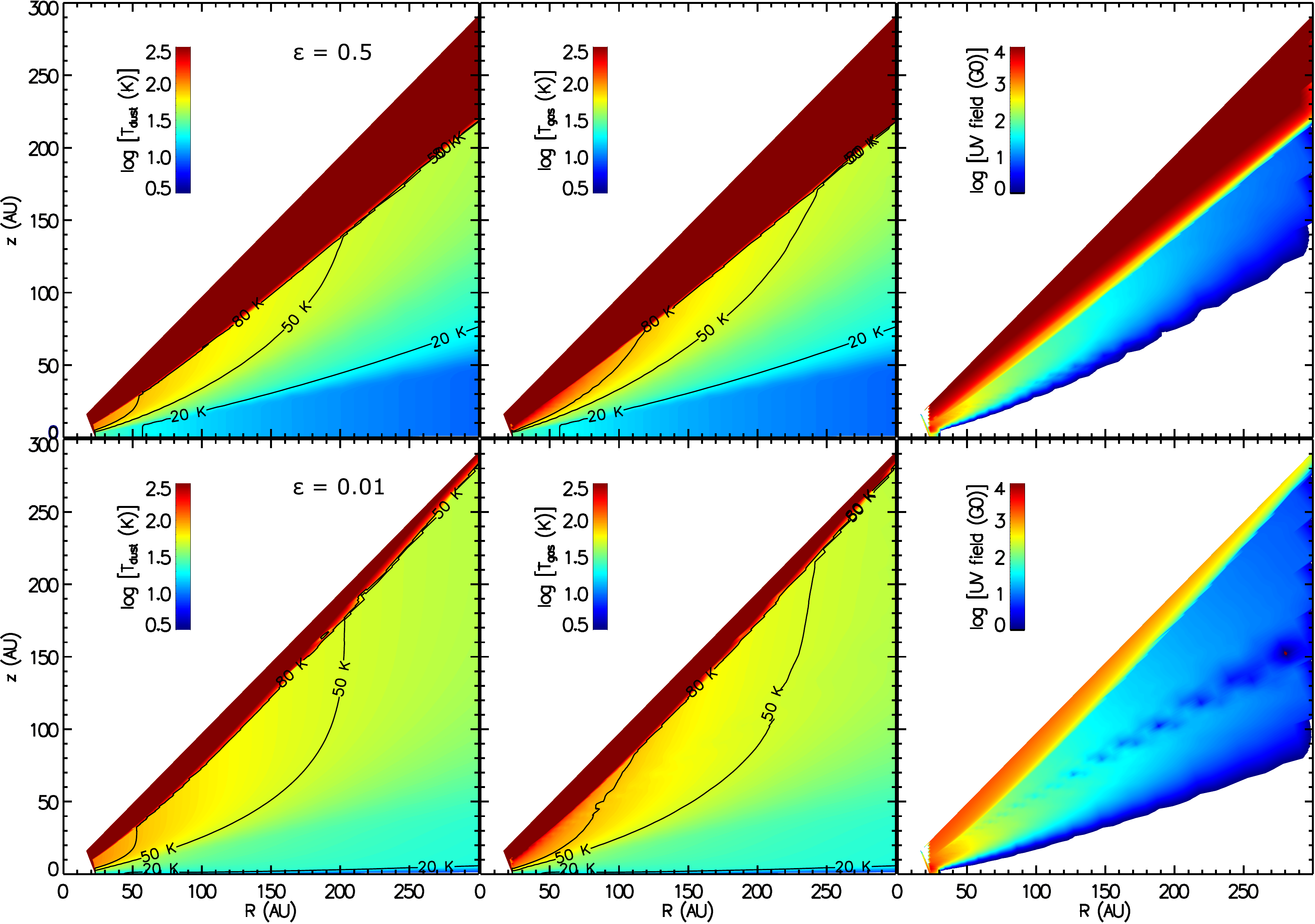}}
\caption{GM Aur disk structures. {\it Upper row:} dust and gas temperature and UV flux structures for $\epsilon_{small}$=0.5 (50\% less dust). {\it Lower row:} dust and gas temperature and UV flux structures for $\epsilon_{small}$=0.01 (99\% less dust). The UV radiation field is given in units of G0, or 1 Habing unit, equal to 1.6$\times$10$^{-3}$ erg s$^{-1}$cm$^{-2}$. \label{gmtstruct}}
\end{figure*}

The effect on the dust temperature structure of varying the dust depletion factor is significant. A smaller dust opacity in the upper layers causes the stellar radiation to be deposited at lower disk altitudes, cooling the uppermost disk layer and warming the lower disk layers (although not the midplane itself). In the leftmost panels of Figures \ref{dmtstruct} and \ref{gmtstruct}, by volume 30\% of the disk is colder than 20 K in the less depleted, $\epsilon_{small}$=0.5 model compared with $<$2\% of the more depleted, $\epsilon_{small}$=0.01 model. The increased dust depletion from the upper layers, manifests in the SEDs as a lower overall flux and a bluer slope near 100 $\mu$m, seen in Figure \ref{allseds}. In this figure, a range in dust depletion from 50\% to 99\% relative to the standard dust/gas ratio fits similarly well within the error bars, with a formal best-fitting value of $\epsilon_{small}$=0.05$^{+0.05}_{-0.04}$, or a 95\% depletion, for both GM Aur and DM Tau. The uncertainty comes from the fact that DM Tau and GM Aur are transitional (i.e. their millimeter emission displays evidence of an inner clearing close to the star), and the frontally illuminated inner edge of the dusty disk dominates the mid-infrared emission and into the far-infrared regime. Consequently, the impact on their SEDs from varying the degree of dust depletion is less pronounced than for full disks. We are ultimately able to put firmer constraints on the value for $\epsilon_{small}$ through the comparison with the HD line flux below. 

Heating by UV radiation also changes the gas temperature structure substantially, as seen in the upper left and center panels of Figures \ref{dmtstruct} and \ref{gmtstruct}, which show the dust and gas temperatures respectively. However, the increase in temperature occurs mainly in the disk upper layers above the 50 K isotherm. This is because the UV radiation field is attenuated to interstellar values just below the 50 K isotherm, as seen in the upper right panel of Figures \ref{dmtstruct} and \ref{gmtstruct}, so the gas is not heated efficiently at lower disk altitudes.  As discussed in \citet{bergin+13}, the HD emission line strength depends both on the temperature of that gas, in addition to the total mass in HD, as warmer gas populates the J=1 upper state more relative to the J=0 level, compared with cooler gas. The fractional populations are compared in the left column of Appendix Figures \ref{dmfpop} and \ref{gmfpop}, in addition to the total gas density and density of HD in the J=1 level.  For the less dust depleted model, the J=1 density is clearly bimodal, with a strong contribution in the midplane, where the total gas density is strongest, and a spur continuing into the disk upper layers, where the fractional population is highest.  This pattern is similar to that found for TW Hya by \citet{bergin+13}.  In the more depleted model, however, the J=1 level is populated closer to the midplane, so the J=1 density is more reflective of the entire disk density structure than it is in the less depleted model.

As a result, the line emission is an order of magnitude stronger in the most depleted model compared with the least depleted. In Figure \ref{masses_final_fig}, we compare the integrated line flux as a function of gas mass for Case 1 with $T_{gas}>T_{dust}$ and $T_{gas}=T_{dust}$. The higher gas temperature produces only 4 to 6\% increase in the line flux, so we do not consider further the gas temperature separately. The HD line flux provides further constraints on both the value of $\epsilon_{small}$ and the gas mass. Within the observational uncertainties, the new best-fitting value of the dust depletion parameter $\epsilon_{small}$ is 0.03$\pm$0.01 and 0.08$\pm$0.02 for DM Tau and GM Aur, respectively. The new best-fitting value corresponds to a decrease in mass for DM Tau and an increase in mass for GM Aur. However, the difference in line flux between the Case 1 models is due mainly to the distribution of dust and the temperature structure rather than the mass.


\subsection{Case 2: Effect of varying the surface density on HD line emission}

To test directly the impact of gas mass on the line flux, we varied the surface density by increasing $\alpha$ and enhancing the dust/gas mass ratio at the midplane as described above and compared the output line fluxes with the data and Case 1 models in Figure \ref{masses_final_fig}. Changing the dust/gas ratio at the midplane produces some variation in the temperature structure, but it is small compared with the effect of changing the dust/gas ratio in the upper layers of the disk. 

The HD line flux varies directly with mass, as we expected. For DM Tau and GM Aur, the range of masses consistent with the observed line is broad: half an order of magnitude and a full order of magnitude respectively, with d/g at the midplane $\sim$0.5 to 0.07. However the line flux is more sensitive to changes in gas mass when the upper layers are more dust depleted, as seen by comparing the $\epsilon_{small}$=0.02 Case 2 family of models with those for $\epsilon_{small}$=0.05 for DM Tau in Figure \ref{masses_final_fig}. This is because a larger fraction of the HD is in the J=1 state for lower values of $\epsilon_{small}$, as explained above. The other disks in this sample have lower values of $\epsilon_{small}$, more consistent with the median value for Taurus \citep[$\epsilon_{small}$=0.01][]{furlan+06}. Thus the J=1-0 transition can be a good tracer of mass for a majority of disks, provided that the temperature structure is well-constrained.

\subsection{Gas masses}
\label{gmass}

\begin{figure*}
\centering{\includegraphics[angle=0, scale=0.6]{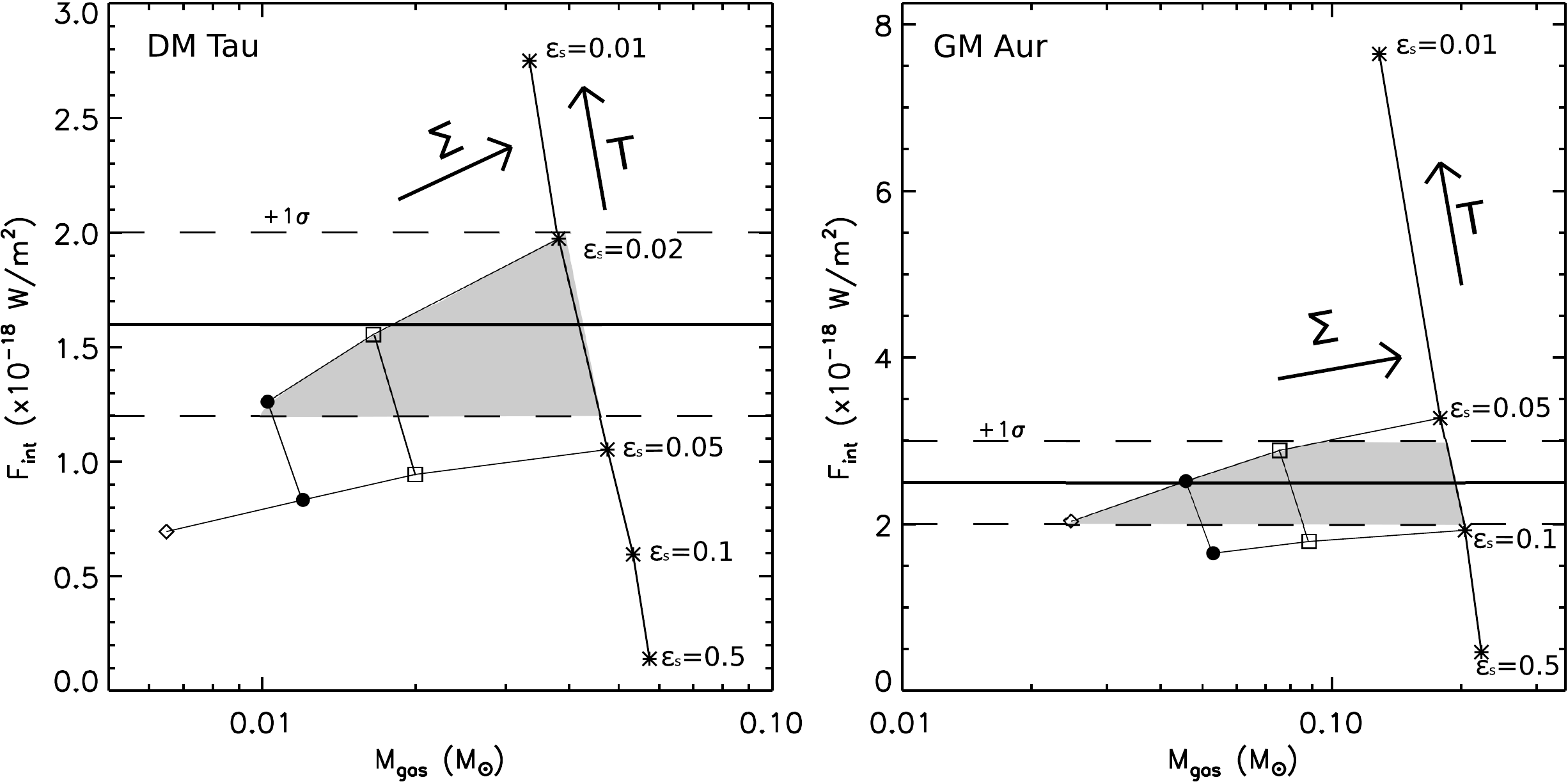}}
\caption{Integrated line flux as a function of total gas mass in a grid of different upper layer temperature structures and disk surface densities for DM Tau (left) and GM Aur (right). The range of disk gas masses permitted by the SED fitting {\it and} the HD J=1-0 line flux is indicated by the grey fill.  Arrows indicate the direction in the grid along which the temperature structure and surface density vary the most. Different symbols denote models with decreasing surface density and increase midplane dust concentration: $\alpha$=0.001 and $\epsilon_{big}$ defined by vertical mass conservation (asterisks), $\alpha$=0.002 and $\epsilon_{big}$=25 (open squares), $\alpha$=0.003 and $\epsilon_{big}$=50 (filled circles), and $\alpha$=0.005 and $\epsilon_{big}$=75  (open diamonds).  UV gas-heating provides an additional 4 - 6\% increase in the line emission. The best-fitting mass ranges are 1.9$\times$10$^{-2}$-4.3$\times$10$^{-2}$ $M_{\odot}$ for DM Tau, and 4.5$\times$10$^{-2}$-19.5$\times$10$^{-2}$ $M_{\odot}$ for GM Aur, although values of 1.0-4.7$\times10^{-2}$ and 2.5-20.4$\times10^{-2}$, respectively, are permitted within the $\pm$1$\sigma$ uncertainties. \label{masses_final_fig}}
\end{figure*}

Considering the full range of Case 1 and 2 models that fit both the SED and 1$\sigma$ uncertainties on the HD line flux, there are a range of masses for the disks with HD detections: 1.0-4.7$\times10^{-2}$ for DM Tau and 2.5-20.4$\times10^{-2}$ for GM Aur. Although the mass for GM Aur is large, the disk is still marginally stable, with Toomre Q $\sim$1.3 at 300 AU. The masses corresponding to the model grid are shown in Figure \ref{masses_final_fig}. The gray polygon indicates the parameter space that fits both the dust SED and the gas HD line. The maximum values for $\alpha$ and the increased dust/gas ratio were not physically motivated; i.e. to see how small a mass we could obtain, we increased these parameters until the models no longer fit the SED without considering whether such large values were realistic.  Even with these extreme values, DM Tau's mass is greater than or equal to the minimum mass solar nebula. The mass of GM Aur is larger by at least a factor of two, even with the smallest surface density.

The masses we derive for these two disks may be underestimated due to our model assumption of total dust and gas depletion in their gaps. In contrast to this assumption, recent ALMA observations have detected gas inside of the dust gap for several disks in the transitional disk class with depletions of two to three orders of magnitude relative to the gap edge \citep{vandermarel+16a}. If this is also the case for DM Tau and GM Aur, then in this work we have underestimated their disk gas masses. 

Compared with the mass published by \citet{bergin+13} for TW Hya, 0.06 M$_{\odot}$, DM Tau has a mass that is 20\% lower. However, when corrected for the difference in distance DM Tau has a 64\% brighter line detection. This discrepancy is likely due to a difference in the dust distribution in the upper layers, which has a large effect on the line flux as demonstrated by Figure \ref{masses_final_fig}. If TW Hya is less settled than DM Tau, it would have a cooler mid-altitude temperature structure and therefore a lower line flux for the same gas mass.

\begin{deluxetable}{cccc}
\tabletypesize{\scriptsize}   
\tablewidth{0pt}
\tablecaption{Best-fitting dust and gas masses, Cases 1 and 2}
\tablehead{\colhead{Value} & \colhead{DM Tau} & \colhead{GM Aur} & \colhead{Ref.}}
\startdata
M$_{dust}$ (M$_{\odot}$) & 2.9$\times$10$^{-4}$ & 1.25$\times$10$^{-3}$ &  \\
M$_{gas}$ (M$_{\odot}$) & (1.0-4.7)$\times10^{-2}$ & (2.5-20.4)$\times10^{-2}$ &  \\
$\epsilon_{small}$ &	0.02-0.04	&	0.06-0.1	\\
$\alpha$  & 0.001-0.005 & 0.001-0.005 \\
$\epsilon_{big}$ & 12.4 - 50	&	11.4 - 75 & \\
\hline
M$_{dust,lit.}$ (M$_{\odot}$)& 1.5$\times$10$^{-4}$  &			& 2			\\					   
M$_{gas,lit.}$ (M$_{\odot}$) & 1.4$\times$10$^{-3}$ &  $\leq$0.35$\times$10$^{-3}$ & 1 \\
					   & 	9.0$\times$10$^{-3}$  & 			& 2 
\enddata
\tablecomments{The dust/gas mass ratio for each region (disk upper layers and midplane) is $\epsilon \times 0.0065$. 
References: (1) \citet{dutrey+96}, (2) \citet{williams+14a}}
\label{masstab}
\end{deluxetable}

\subsection{Non-detections}
\label{nodet}
For the disks in which HD was not detected, we conducted a Case 1 model analysis in the same way as for DM Tau and GM Aur, testing the full $\epsilon_{small}$ and $\alpha$ parameter space. The HD line flux for the best fit to the SED was calculated and compared with the upper limit on the observed line flux listed in Table \ref{fluxtab}. The best-fitting SED models for two of these disks produce HD line fluxes that are formally just above the observed 3$\sigma$ upper limits to their integrated flux. Their peak central fluxes are still below the 3$\sigma$ level of the continuum: VZ Cha (2.5 $\times 10^{-18}$ W m$^{-2}$) and AA Tau (2.5$\times 10^{-18}$ W m$^{-2}$). This puts a firm lower limit on $\epsilon_{small}$, as a lower value would produce stronger HD line emission. Another way to view the limit on $\epsilon_{small}$ is as an upper limit to the gas/dust ratio in the disk upper layers. For VZ Cha, the gas/dust mass ratio can be no larger than 153,800, while for AA Tau it must be lower than 15,380.

Since the maximum value of $\epsilon_{small}$ is better constrained for these two disks than for DM Tau and GM Aur (as they are thought to be full, non-transitional disks), and the value of the dust/gas ratio in the midplane of the Case I models is a lower limit for our parameter space, the total masses for these disks listed in Table \ref{modtab} are true upper limits. The other two disks have integrated line fluxes well below their 3$\sigma$ upper limits from Table \ref{fluxtab}: FZ Tau (1.8$\times 10^{-18}$ W m$^{-2}$) and LkCa 15 (6.7$\times 10^{-19}$ W m$^{-2}$). For these disks, the derived total mass is an upper limit as well, simply due to the limits on $\epsilon_{small}$ provided by the SED fitting; however the HD line flux itself imposes no additional constraints.

\section{Discussion}
\label{discussion}
\subsection{Comparing CO and HD gas masses: carbon depletion?}
\label{cohd}
The range of gas masses found here results directly from the strong dependence of the HD line emission on both the gas temperature and density structures. Masses derived from CO exhibit similar dependencies, with the additional uncertainty of freeze-out and chemical sequestration of carbon into other, more complex species \citep{bruderer+12,bruderer+13,favre+13, bergin+14a, reboussin+15, kama+16a, kama+16b}. Gas masses can be further underestimated if isotope-selective photodissociation is not properly taken into account \citep{miotello+14a}. Contrasting our gas masses with those derived from CO could independently confirm the chemical depletion of carbon in these disks.

For both DM Tau and GM Aur, gas masses of 1.4$\times$10$^{-3}$ and $\leq$0.35$\times$10$^{-3}$, respectively, were derived from CO measurements by \citet{dutrey+96}. This work did not account for either freeze-out, photodissociation, or chemical depletion, and the fact that their gas masses are at least an order of magnitude less than what we measure with HD is consistent with the expectation that the effects of freeze-out and photodissociation biases the CO-derived gas masses to lower values.  A more recent work, \citet{williams+14a}, calculates a disk gas mass from CO using a generic model grid that takes these two effects into account and compares this grid with a sample of disks that includes DM Tau only. Although \citet{williams+14a} take into account optical depth effects by using CO isotopologs and parameterizing freeze-out and photodissociation, in the locations where CO should be present in the gas phase they use a single value of CO/H$_2$=1$\times$10$^{-4}$ to determine the final gas mass. The mass that they derive from CO is consistent with our lower limit (0.9$\times$10$^{-2}$ vs. 1.0$\times$10$^{-2}$ M$_{\odot}$) but 5 times lower than our upper limit of 4.7$\times$10$^{-2}$ M$_{\odot}$. This suggests that even after photodissociation and freeze-out are accounted for, the CO in DM Tau may be chemically depleted by up to a factor of 5.

While we cannot entirely rule out their CO derived gas mass, it is worth noting a few caveats. First, as discussed in Section \ref{gmass}, our lower limit to the mass requires a dust/gas enhancement at the midplane that is not physically motivated.  Second, neither the gas masses derived from CO nor HD account completely for the presence of depleted gas and dust in the gaps or inner clearings of these transitional disks. Our work assumes a totally depleted gap, resulting in an under-estimate of the disk mass, as described in Section \ref{gmass}. The CO gas masses were derived with the assumption of a full disk, i.e. a disk without gaps or clearings, and therefore overestimates the gas mass generally. Some secondary effects would also apply: the warmer midplane gas temperatures in the gap should prevent CO freeze-out there, while the additional dust depletion and greater UV penetration might allow more photodissociation. If those two effects cancel each other, then overall the modeled CO gas mass would be lower, while our mass should be higher.

While GM Aur was not modeled by \citet{williams+14a}, their gas mass for DM Tau was a factor of six greater than the mass found by \citet{dutrey+96}, which did not include freeze-out or photodissociation. Extrapolating the gas mass found for GM Aur by \citet{dutrey+96} by the same factor results in a mass of 2$\times$10$^{-3}$, which would yield a carbon depletion on the order of one to two orders of magnitude. This suggests that GM Aur could be more strongly carbon depleted than DM Tau.

\subsection{Breaking the degeneracies between temperature structure and total gas mass}

In this work, we used a basic dust composition of silicates and graphite with a range of dust/gas ratios and showed how the disk temperature structure, gas mass could vary to produce different fits to the HD line. To constrain further the disk mass distribution and compare it against similar theoretical studies, it is necessary to combine these HD measurements with additional gas and dust data at sufficiently high spatial or spectral resolution to break the current degeneracies between the temperature structure and total gas mass. Since the gas and dust are still coupled at the disk height from which most of the HD J=1 level emits, accurate knowledge of the temperature structure requires understanding the distribution of dust in the upper layers and disk midplane.

The vertical temperature structure of disk depends on the degree of dust settling and the detailed dust composition, both of which can be better constrained by continuum spectral data, i.e. SPIRE observations, than they can be by continuum photometry as we have done here. An analysis of the detailed shape of the SED between 100 and 300 $\mu$m will allow a more precise determination of $\epsilon_{small}$ and the dust properties (Espaillat et al., in prep). Another way to probe specifically the vertical gas temperature structure in the region of the disk where HD emits is through line observations of CO. Simultaneous modeling of the CO ladder in the far-infrared has constrained the temperature structure in the upper layers of several bright disks \citep{fedele+13a,fedele+16}. At longer wavelengths, \citet{rosenfeld+13a} find evidence for the vertical location of the CO freeze-out zone (typically T$_{dust}\sim$20K) and the location of the surface where $^{12}$CO becomes optically thick. These type of measurements, in conjunction with less optically thick CO isotopologues, could be used to pin down the temperature structure at different depths in the disk \citep[][Miotello et al., in prep]{cleeves+15a, kama+16a, kama+16b}. Additionally, observations of the J=6-5 transition of $^{13}$CO, which has a similar upper state energy to the HD J=1-0 transition, can resolve the temperature structure over a similar region to the HD transition \citep{schwarz+16a}. Once the temperature structure is determined, a more precise gas mass can be determined from Figure \ref{masses_final_fig}.

Another way to constrain the mass in Figure \ref{masses_final_fig} is to combine the HD measurements with constraints on the midplane temperature structure from resolved observations of the CO snowline. Varying the dust/gas ratio changes the midplane dust temperature. For example, in the case of GM Aur models in which the vertical dust mass has been conserved have the CO snowline ($T$=20K) at 51 to 53 AU. These are the models with the largest disk gas mass. At the other end of the mass range, with models that include a dust/gas ratio enhanced by some additional effect (e.g. radial drift), the CO snowline lies at 44 AU. ALMA has provided resolved observations of the CO snowline already for some disks \citep{qi+13}; similar observations would put strong constraints on the midplane dust/gas ratio and hence the total gas mass.

\section{Conclusions}
\label{conclusions}

In this work we have detected the HD J=1-0 transition in two TTauri disks, DM Tau and GM Aur, out of a sample of six.  Using a combination of SED and line emission models, we determine the ranges of mass permitted to these disks, given the uncertainties in the temperature structure and dust/gas ratio from the observational uncertainties. 
\begin{itemize}
\item DM Tau and GM Aur are found to have masses of 1.0-4.7$\times10^{-2}$ and 2.5-20.4$\times10^{-2}$, respectively. 
\item For the disks in which HD is not detected, the non-detections provide strict limits on the gas/dust ratio in the disk upper layers of $<$153,800 and $<$15,380 for VZ Cha and AA Tau, respectively, while limits could be established for LkCa15 and FZ Tau only on the gas/dust ratio from SED fitting.
\item Comparison of our HD gas masses and those derived from CO suggests that GM Aur shows gas phase carbon depletion of up to two orders of magnitude, while DM may show depletion of up to a factor of five. 

\end{itemize}

Going forward, a combination of physically motivated models with disk chemistry that fit HD, CO, and the dust SED will be able to disentangle the disk temperature and density structures, while determining the degree of carbon depletion. Observations of the HD J=1-0 line are a particularly useful way to constrain the disk mass; future facilities like SPICA will hopefully make this transition available for more sources. 

\acknowledgments
This work is based on observations made with Herschel, a European Space Agency Cornerstone Mission with significant participation by NASA. Support for this work was provided by NASA through an award issued by JPL/Caltech. We thank Davide Fedele for useful discussions regarding the data reduction and line flux determinations. We would also like to thank the anonymous referee for constructive suggestions.


\clearpage

\appendix

\section{Fractional populations of HD in the J=1 state}
\begin{figure*}[b]
\centering{\includegraphics[angle=0, scale=0.4]{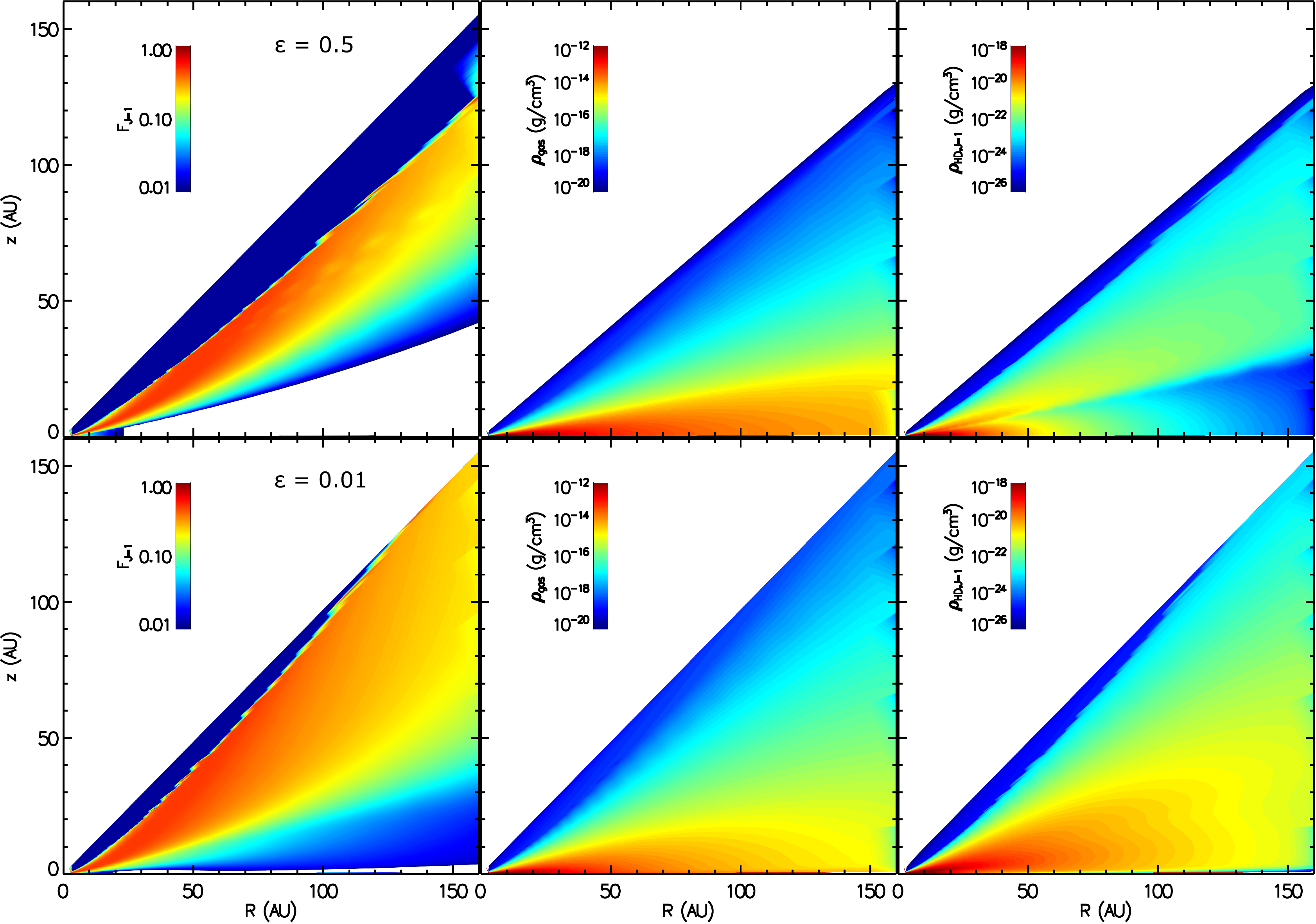}}
\caption{DM Tau disk structures. {\it Upper row:} fractional HD population in the J=1 level (left), gas density structure (center), and HD J=1 density structure (right) for $\epsilon_{small}$=0.5 (50\% less dust). {\it Lower row:} fractional HD population in the J=1 level (left), gas density structure (center), and HD J=1 density structure (right) for $\epsilon_{small}$=0.01 (99\% less dust). \label{dmfpop}}
\end{figure*}
\begin{figure*}[b]
\centering{\includegraphics[angle=0, scale=0.4]{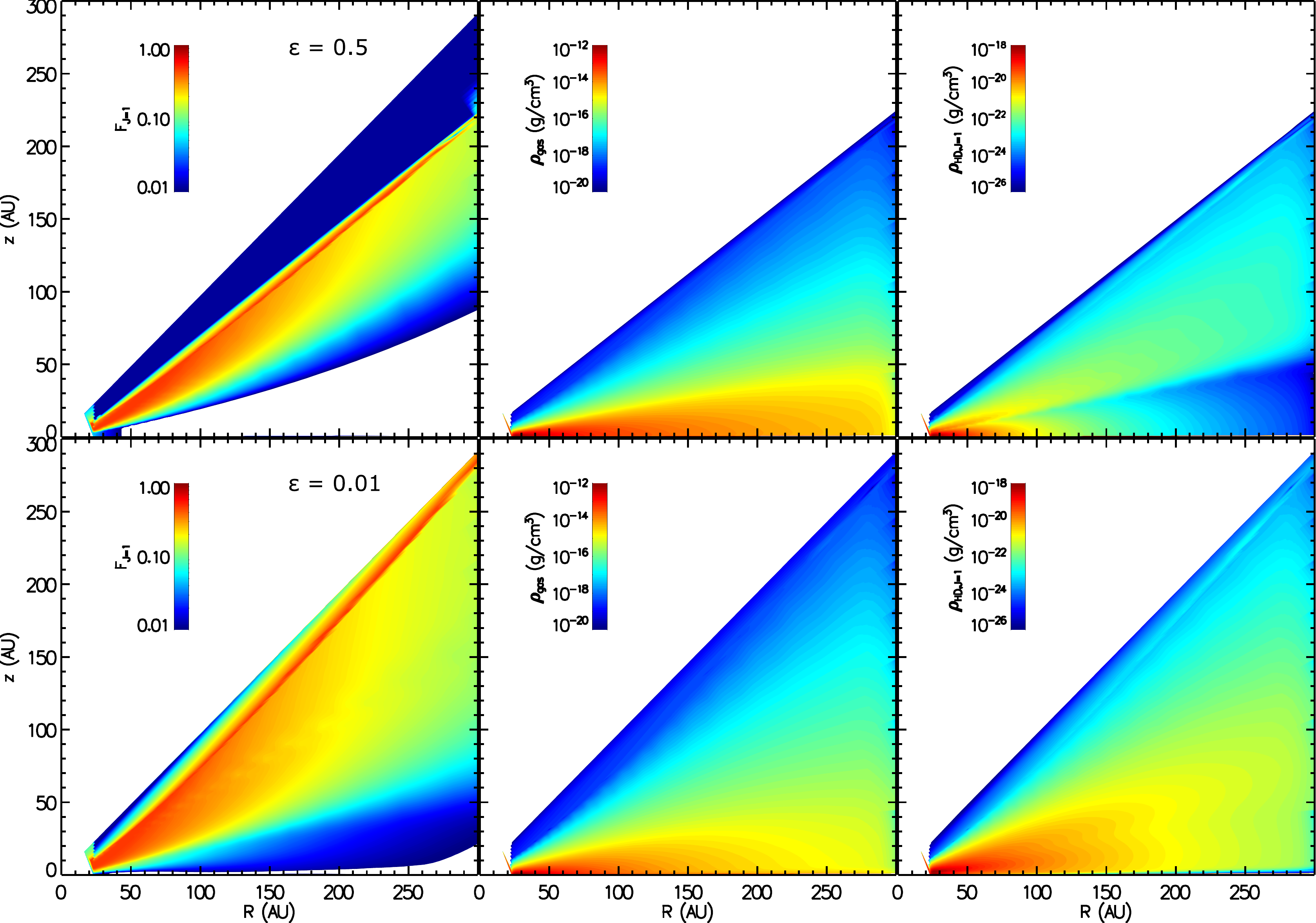}}
\caption{GM Aur disk structures. {\it Upper row:} fractional HD population in the J=1 level (left), gas density structure (center), and HD J=1 density structure (right) for $\epsilon_{small}$=0.5 (50\% less dust). {\it Lower row:} fractional HD population in the J=1 level (left), gas density structure (center), and HD J=1 density structure (right) for $\epsilon_{small}$=0.01 (99\% less dust). \label{gmfpop}}
\end{figure*}


\end{document}